\pdfoutput=1

\documentclass[onecolumn]{emulateapj}

\usepackage{amsmath}
\usepackage{graphicx}
\usepackage{xcolor}
\usepackage[colorlinks]{hyperref}
\hypersetup{
    colorlinks,	
    citecolor=blue,
}

\usepackage{amsfonts}
\usepackage{amsmath}
\usepackage{amssymb}
\usepackage{bm}
\usepackage{dcolumn}
\usepackage{graphicx}
\usepackage{graphics}
\usepackage[latin1]{inputenc}
\usepackage{latexsym}
\usepackage{rotating}
\usepackage{xspace} 
\usepackage{mathrsfs}
\usepackage{hyperref}
\usepackage{epstopdf}
\usepackage{gensymb}

\shorttitle{Public Release of \textsc{RELXILL\_NK}}
\shortauthors{Abdikamalov et al.}

\begin{document}

\title{Public Release of \textsc{relxill\_nk}: A Relativistic Reflection Model for Testing Einstein's Gravity}

\author{Askar~B.~Abdikamalov\altaffilmark{1}, Dimitry~Ayzenberg\altaffilmark{1}, Cosimo~Bambi\altaffilmark{1,\dag}, Thomas~Dauser\altaffilmark{2}, Javier~A.~Garc\'ia\altaffilmark{3,2}, and Sourabh~Nampalliwar\altaffilmark{4}}

\altaffiltext{1}{Center for Field Theory and Particle Physics and Department of Physics, 
Fudan University, 200438 Shanghai, China. \email[\dag E-mail: ]{bambi@fudan.edu.cn}}
\altaffiltext{2}{Remeis Observatory \& ECAP, Universit\"at Erlangen-N\"urnberg, 96049 Bamberg, Germany}
\altaffiltext{3}{Cahill Center for Astronomy and Astrophysics, California Institute of Technology, Pasadena, CA 91125, USA}
\altaffiltext{4}{Theoretical Astrophysics, Eberhard-Karls Universit\"at T\"ubingen, 72076 T\"ubingen, Germany}

\date{\today}

\begin{abstract} 

We present the public release version of \textsc{relxill\_nk}, an X-ray reflection model for testing the Kerr hypothesis and general relativity. This model extends the \textsc{relxill} model that assumes the black hole spacetime is described by the Kerr metric. We also present \textsc{relxilllp\_nk}, the first non-Kerr X-ray reflection model with a lamppost corona configuration, as well as all other models available in the full \textsc{relxill\_nk} package. In all models the relevant relativistic effects are calculated through a general relativistic ray-tracing code that can be applied to any well-behaved, stationary, axisymmetric, and asymptotically flat black hole spacetime. We show that the numerical error introduced by using a ray-tracing code is not significant as compared with the observational error present in current X-ray reflection spectrum observations. In addition, we present the reflection spectrum for the Johannsen metric as calculated by \textsc{relxill\_nk}.

\end{abstract}

\keywords{accretion, accretion disks --- black hole physics --- gravitation}

\section{Introduction}

Observations of black hole (BH) accretion processes are one of the few available probes of the strong-field regime of gravity in the vicinity of black holes [see e.g.~\citet{2017RvMP...89b5001B, 2018AnP...53000430B} for a review]. These observations, in principle, allow for the determination of the properties of the BH spacetime, such as the BH mass and BH spin angular momentum. Currently, the two well-established approaches to study these observations are the continuum-fitting method and X-ray reflection spectroscopy. These methods have been used to estimate the spins of about a dozen stellar-mass BHs and about twenty supermassive BHs~\citep{2018AnP...53000430B}. A third approach is the study of quasi-periodic oscillations in the X-ray power density spectrum. However, the exact nature of these oscillations is still not well understood.

In addition to determining the properties of BHs, these observations of BHs with accretion disks can, in principle, be used to test the \textit{Kerr hypothesis}. The Kerr hypothesis states that the correct description for all isolated, stationary, and axisymmetric astrophysical (uncharged) BHs is the Kerr metric~\citep{1975PhRvL..34..905R, 1967PhRv..164.1776I, 1968CMaPh...8..245I, 1971PhRvL..26.1344H, 1972CMaPh..25..152H, 1971PhRvL..26..331C}. The Kerr metric is completely determined by two parameters: the BH mass and the BH spin angular momentum. The Kerr hypothesis holds in general relativity (GR) and in some modified gravity theories~\citep{Psaltis:2007cw}, but there are some theories in which it does not [e.g. Chern-Simons gravity~\citep{Alexander:2009tp}]. BHs within these theories are not described by the Kerr metric, and thus, BH accretion disk observations can, in principle, test GR and place constraints on modified gravity theories in which the Kerr hypothesis is violated.

In this work we focus on the X-ray reflection spectroscopy method used to study the properties of BHs with accretion disks. In particular, we are interested in the prospects of using observations of the X-ray reflection spectrum to test the Kerr hypothesis. Currently the most advanced model for calculation of the reflection spectrum is \textsc{relxill}~\citep{doi:10.1093/mnras/sts710, 0004-637X-782-2-76}. However, \textsc{relxill} is limited to the reflection spectrum of accretion disks around Kerr BHs. With such a model it is still possible to test the Kerr hypothesis, as any significant deviations away from Kerr would significantly modify the spectrum. However, it is more difficult to do so and, in particular, placing constraints on modified gravity theories is not possible. The latter requires a X-ray reflection spectrum model that can incorporate a wide range of BH solutions.

In this paper, we present the public release version of \textsc{relxill\_nk}{\footnote{\textsc{relxill\_nk} package available at \href{http://www.physics.fudan.edu.cn/tps/people/bambi/Site/RELXILL_NK.html}{http://www.physics.fudan.edu.cn/tps/people/bambi/Site/ RELXILL\_NK.html} and \href{http://www.tat.physik.uni-tuebingen.de/~nampalliwar/relxill_nk/}{http://www.tat.physik.uni-tuebingen.de/$\sim$nampalliwar/relxill\_nk/}. For support contact \href{mailto:relxill_nk@fudan.edu.cn}{relxill\_nk@fudan.edu.cn}.}}~\citep{Bambi:2016sac}, an extension of the relativistic X-ray reflection model \textsc{relxill}~\citep{doi:10.1093/mnras/sts710, 0004-637X-782-2-76} to include any well-behaved, stationary, axisymmetric, and asymptotically flat black hole metric, allowing for tests of the Kerr black hole hypothesis. As in \textsc{relxill}, we use the formalism of the Cunningham transfer function for thin accretion disks~\citep{1975ApJ...202..788C, 1995CoPhC..88..109S, 2010MNRAS.409.1534D} to compute all of the relativistic effects on the emission from the disk. However, since not every metric is necessarily separable like the Kerr metric, to keep our code more general we do not assume separability and the task of computing the transfer function cannot be reduced to quadrature as in the Kerr case. Instead, we use a general relativistic ray-tracing code to solve the null geodesic equations of motion for photons emitted from the disk and seen by a distant observer. Using such a method increases the numerical error, however, we show that the numerical error introduced by our methodology is well below the observational error present in current X-ray reflection spectrum observations and thus is not a cause for concern at the moment. The base \textsc{relxill\_nk} model has already been used to analyze the X-ray reflection spectra of a number of BHs and place constraints on some non-Kerr metrics~\citep{Cao:2017kdq, Zhou:2018bxk, Tripathi:2018bbu, Wang-Ji:2018ssh, Xu:2018lom, Choudhury:2018zmf, Tripathi:2018lhx, Tripathi:2019bya}.

Additionally, we present the new model \textsc{relxilllp\_nk}, which extends \textsc{relxilllp}~\citep{doi:10.1093/mnras/sts710} where, rather than assuming some emission profile from the disk, the emission profile is determined from the impinging radiation profile due to a isotropically-emitting point source corona at some height along the spin axis of the BH. This is referred to as the lamppost geometry corona model and naturally explains the steep emissivity observed in the reflection spectrum~\citep{1991A&A...247...25M, 1996MNRAS.282L..53M, 2002A&A...383L..23M, 2011A&A...533L...3D, 2011MNRAS.414.1269W, 2012MNRAS.422.1914D}. We use the same general relativistic ray-tracing code as in the standard \textsc{relxill\_nk} model and solve the null geodesic equations of motion for photons traveling from the corona down to the disk. We also show that using our ray-tracing method does not significantly increase the numerical error present in the model.

This paper is organized as follows. Section~\ref{sec:xray} describes the basics of X-ray reflection spectroscopy. Section~\ref{sec:relxill_nk} explains how the reflection spectrum is calculated in \textsc{relxill\_nk}. Section~\ref{sec:comp} shows the accuracy of \textsc{relxill\_nk} as compared with \textsc{relxill} in the Kerr background. Section~\ref{sec:mods} summarizes the available models in the \textsc{relxill\_nk} package and shows the effect of a non-Kerr background on the reflection spectrum. Section~\ref{sec:concs} concludes by summarizing and discussing possible future improvements to \textsc{relxill\_nk}.

\section{X-ray Reflection Spectroscopy}
\label{sec:xray}

We model the BH-disk system using the standard disk-corona model~\citep{2017bhlt.book.....B, 2018AnP...53000430B}, in which the BH is surrounded by a geometrically-thin and optically-thick accretion disk and there is a nearby cloud of hotter gas termed a ``corona". The disk is assumed to be in the equatorial plane of the BH and extends from some outer radius $R_{\text{out}}$ to an inner radius $R_{\text{in}}$, which is generally assumed to be at or near the innermost stable circular orbit (ISCO) radius of the BH. The emission of the disk is locally a blackbody and becomes a multi-temperature blackbody when integrated radially; this is known as the \textit{thermal component} of the total BH spectrum. Locally the temperature depends on the mass of the BH, the accretion rate, and the distance from the BH. With an accretion rate of about~$10\%$ of the Eddington rate, the thermal spectrum of the inner part of the disk is in the soft X-ray band (0.1-1 keV) for stellar-mass BHs and in the optical/UV band (1-10 eV) for supermassive BHs. Note that currently our model does not include the thermal emission from the disk.

The corona is modeled as a significantly hotter ($\sim100$ keV), usually optically thin, cloud somewhere in the vicinity of the BH and disk~\citep{2017bhlt.book.....B, 2018AnP...53000430B}. The most common geometries for the corona are a point or spherical source along the spin axis of the BH to represent the base of some jet or a layer above and below the accretion disk to represent some additional atmosphere, but the exact morphology is not yet known.

\begin{figure}
\begin{center}
\includegraphics[width=8.5cm]{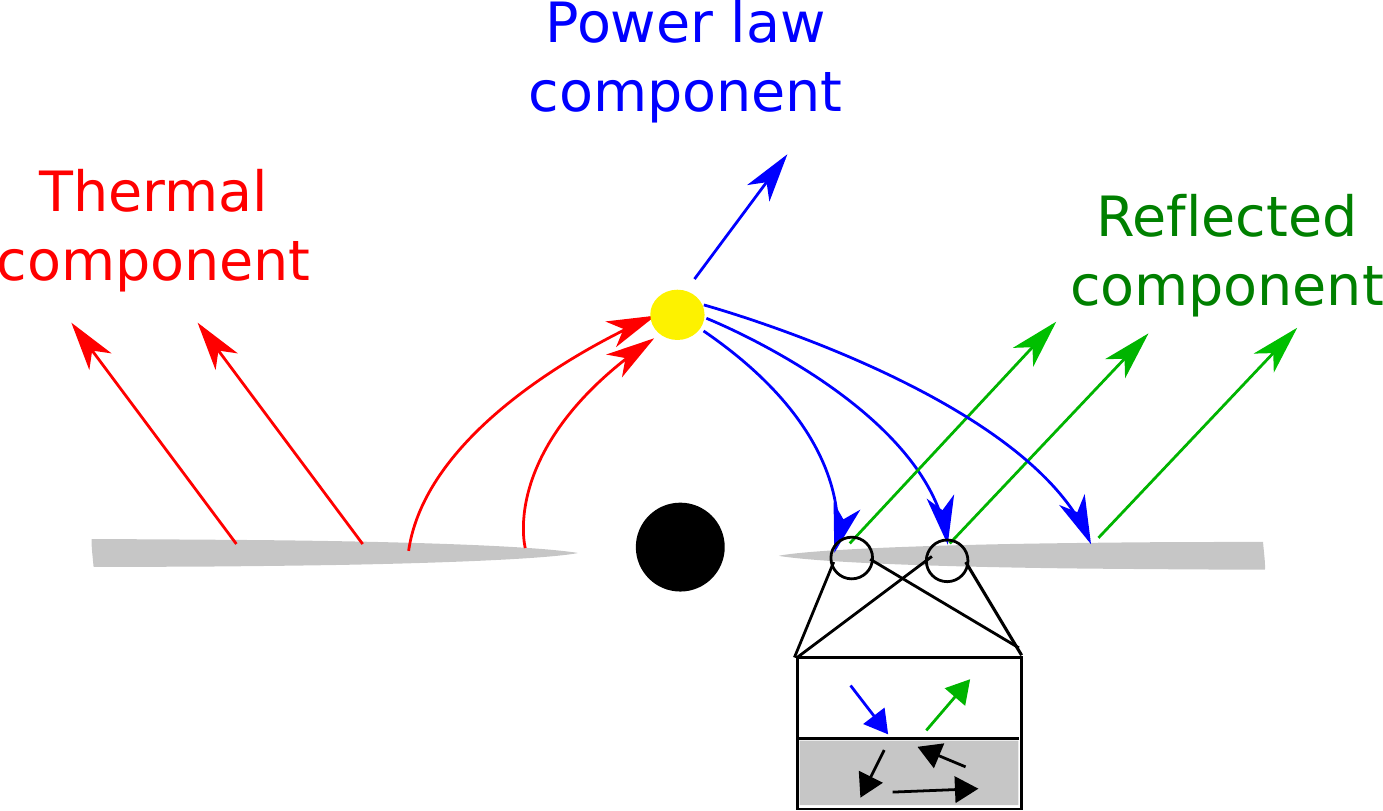}
\end{center}
\vspace{0.3cm}
\caption{\label{fig:diskcorona} Sketch of the disk-corona model and reflection process.}
\vspace{0.3cm}
\end{figure}

The reflection spectrum is produced by interaction between the accretion disk and the corona. The thermal photons produced by the disk inverse Compton scatter off free electrons in the corona, in turn producing a \textit{power-law component} with a cut-off energy that depends on the temperature of the corona (typically $E_{\text{cut}}\sim30-300$ keV). This power-law component then illuminates the accretion disk and is re-emitted as a \textit{reflection component} that includes fluorescent emission lines~\citep{Garcia:2013oma}. The most prominent feature in the reflection component is usually the K$\alpha$ iron line at 6.4 keV in the case of neutral or weakly-ionized iron up to 6.97 keV for H-like iron ions. A sketch of the disk-corona model and reflection process is shown in Fig.~\ref{fig:diskcorona}.

In the rest-frame of the emitter the K$\alpha$ iron line is a very narrow feature, but becomes broadened and skewed in the observer's frame due to the relativistic effects of the BH spacetime (gravitational redshift, Doppler boosting, light bending)~\citep{2017bhlt.book.....B, Fabian:2000nu, Reynolds:2013qqa, Brenneman:2013oba}. This makes observations of the K$\alpha$ iron line a useful tool for studying the properties of BHs with accretion disks. It is important to note, however, that accurate measurements of BH properties require the study of the whole reflection spectrum and not just the iron line.

Models of the reflection component depend on a number of physical parameters of the BH and the accretion disk. The important accretion disk parameters are the inner edge of the disk $R_{\text{in}}$, the outer edge of the disk $R_{\text{out}}$, the inclination angle of the disk $\iota$,~i.e.~the angle between the observer's line of sight and the angular momentum of the disk, the iron abundance $A_{\text{Fe}}$ in solar units (in current popular models all other elemental abundances are assumed to be solar), the ionization of the disk $\xi$ ($\xi=4\pi F_{x}/n$, where $F_{x}$ is the flux and $n$ is the gas density), and parameters related to the emissivity profile of the disk. The emissivity profile depends on the geometry of the corona, and as that is currently unknown the correct profile is not clear. For arbitrary geometries the emissivity profile can be modeled with a power-law (the intensity on the disk $I\propto1/r^{q}$, where $q$ is the emissivity index) or with a broken power-law ($I\propto1/r^{q_{\text{in}}}$ for $r<R_{\text{br}}$ and $I\propto1/r^{q_{\text{out}}}$ for $r>R_{\text{br}}$, where $q_{\text{in}}$ and $q_{\text{out}}$ are the inner and outer emissivity indices, respectively, and $R_{\text{br}}$ is the breaking radius). The incident spectrum on the disk is assumed to be a power law with index $\Gamma$ and some models include a reflection fraction $R_f$ defined as the ratio of intensity emitted towards the disk from the corona compared to the intensity escaping to infinity. In the case of Kerr BHs the relevant parameter is the dimensionless spin of the BH $a^{*}\equiv|\vec{J}|/M^{2}$, where $\vec{J}$ is the spin angular momentum of the BH and $M$ is the mass of the BH. Note that the mass of the BH does not directly influence the reflection component and that the spin angular momentum of the BH is aligned with the angular momentum of the disk in the BH-disk model we are using. For supermassive BHs it is also usually necessary to include the cosmological redshift $z$.

\section{\textsc{relxill\_nk}}
\label{sec:relxill_nk}

\textsc{relxill} is currently the most advanced model for the calculation of the reflection spectrum of accretion disks around Kerr BHs~\citep{doi:10.1093/mnras/sts710, 0004-637X-782-2-76}. \textsc{relxill} is based on the non-relativistic X-ray reflection code \textsc{xillver}~\citep{2010ApJ...718..695G, Garcia:2013oma} and the relativistic line emission code \textsc{relline}~\citep{2010MNRAS.409.1534D, doi:10.1093/mnras/sts710, Dauser:2014jka}. \textsc{relxill} contains a superior treatment of radiative transfer and Compton redistribution as compared to previous codes, and allows for an angular dependence of the reflected spectrum. By implementing the photoionization routines of the \textsc{xstar} code~\citep{Kallman:2001zz}, which is the most complete modeling code for synthetic photoionized X-ray spectra, \textsc{relxill} also improves the calculation of the ionization balance.

The goal of this work is to extend \textsc{relxill} to allow for the modeling of the reflection spectra of non-Kerr BHs. We name this extension collectively as \textsc{relxill\_nk}~\citep{Bambi:2016sac}. As the atomic physics in the disk does not depend on the properties of the spacetime (assuming the Einstein equivalence principle is not violated), no modification of the \textsc{xillver} portion of \textsc{relxill} is required. The parts of the model that must be modified are those that specifically deal with the relativistic effects (e.g.~gravitational redshift, Doppler boosting, light bending), so we will focus on these and not discuss \textsc{xillver} in detail. \textsc{relxill} models the relativistic effects by using the Cunningham transfer function~\citep{1975ApJ...202..788C, 1995CoPhC..88..109S, 2010MNRAS.409.1534D}. We use the same formalism for \textsc{relxill\_nk}, described in Sections~\ref{sec:disk} and~\ref{sec:transfer}, however a different method of computation must be used to calculate the transfer functions. The Kerr solution admits a third constant of the motion, known as the Carter constant, which in turn makes the equations of motion in Kerr separable. This separability reduces the task of computing the transfer functions to numerically calculating a pair of elliptic integrals. Non-Kerr BH solutions, in contrast, are not necessarily separable and so to make \textsc{relxill\_nk} as general as possible we do not assume separability. To calculate the transfer functions we solve the null geodesic equations that describe the motion of the photons, by using a general relativistic ray-tracing code, as detailed in Section~\ref{sec:num}.

\subsection{Black Hole Spacetime}
\label{sec:BH}

While \textsc{relxill\_nk} allows for the study of BH spacetimes beyond the Kerr solution, we do assume that the spacetime is stationary, axisymmetric, and asymptotically flat. In addition, we exclude any cases where the spacetime contains a naked singularity or pathologies such as a violation of the Lorentzian signature or the existence of closed time-like curves outside the event horizon.

In this work we will focus on the non-Kerr metric proposed by Johannsen~\citep{PhysRevD.88.044002} that is a subset of the larger class of metrics first proposed by Vigeland, Yunes, and Stein~\citep{Vigeland:2011ji}. Note, however, that \textsc{relxill\_nk} has already been used with at least one other metric~\citep{Zhou:2018bxk}. The line element of the Johannsen metric in Boyer-Lindquist (BL) coordinates is given by
\begin{align}
ds^{2}=&-\frac{\tilde\Sigma\left(\Delta-a^{2}A_{2}^{2}\sin^{2}\theta\right)}{B^{2}}dt^{2}+\frac{\tilde\Sigma}{\Delta A_{5}}dr^{2}+\tilde\Sigma d\theta^{2}
\nonumber \\
&+\frac{\left[\left(r^{2}+a^{2}\right)^{2}A_{1}^{2}-a^{2}\Delta\sin^{2}\theta\right]\tilde\Sigma\sin^{2}\theta}{B^{2}}d\phi^{2}
\nonumber \\
&-\frac{2a\left[\left(r^{2}+a^{2}\right)A_{1}A_{2}-\Delta\right]\tilde\Sigma\sin^{2}\theta}{B^{2}}dtd\phi,
\end{align}
where
\begin{align}
&B=\left(r^{2}+a^{2}\right)A_{1}-a^{2}A_{2}\sin^{2}\theta, \quad \tilde\Sigma=\Sigma+f,
\nonumber \\
&\Sigma=r^{2}+a^{2}\cos^{2}\theta, \quad \Delta=r^{2}-2Mr+a^{2},
\end{align}
the four free functions $f$, $A_{1}$, $A_{2}$, and $A_{5}$, are\footnote{The four free functions $f$, $A_{1}$, $A_{2}$, and $A_{5}$, are written as a power series in $M/r$
\begin{align}
&f=\sum_{n=2}^{\infty}\epsilon_{n}\frac{M^{n}}{r^{n-2}}, \quad
A_{1}=1+\sum_{n=0}^{\infty}\alpha_{1n}\left(\frac{M}{r}\right)^{n}, \quad
\nonumber \\
&A_{2}=1+\sum_{n=0}^{\infty}\alpha_{2n}\left(\frac{M}{r}\right)^{n}, \quad
A_{5}=1+\sum_{n=0}^{\infty}\alpha_{5n}\left(\frac{M}{r}\right)^{n}.
\end{align}
In order to correctly recover the asymptotic limit, one must impose $\alpha_{10}=\alpha_{20}=\alpha_{50}=0$. Without loss of generality, we can set $\alpha_{11}=\alpha_{21}=\alpha_{51}=0$ as these can be absorbed into the definition of $M$ and $a$. To satisfy Solar System constraints without fine-tuning, $\epsilon_{2}=\alpha_{12}=0$. Thus, the leading-order deformation parameters that are not tightly constrained by Solar System observations are $\epsilon_{3}$, $\alpha_{13}$, $\alpha_{22}$, and $\alpha_{52}$. See~\cite{PhysRevD.88.044002} for more details.}
\begin{align}
f=&\sum_{n=3}^{\infty}\epsilon_{n}\frac{M^{n}}{r^{n-2}},
\nonumber \\
A_{1}=&1+\sum_{n=3}^{\infty}\alpha_{1n}\left(\frac{M}{r}\right)^{n},
\nonumber \\
A_{2}=&1+\sum_{n=2}^{\infty}\alpha_{2n}\left(\frac{M}{r}\right)^{n},
\nonumber \\
A_{5}=&1+\sum_{n=2}^{\infty}\alpha_{5n}\left(\frac{M}{r}\right)^{n},
\end{align}
and $a=|\vec{J}|/M$ is the spin parameter of the BH.

The Johannsen metric depends on the mass $M$ and spin $a$ of the BH as well as four free functions that encode potential deviations away from the Kerr solution. When $\epsilon_{n}=\alpha_{1n}=\alpha_{2n}=\alpha_{5n}=0$ this metric reduces to the Kerr solution. In this work, for simplicity, we will focus on the two cases where only $\alpha_{13}$ or only $\alpha_{22}$ is non-vanishing. Note, these are also the two parameters that have the largest impact on the spacetime~\citep{PhysRevD.88.044002}.

In the Kerr spacetime, the condition for the existence of an event horizon is $a^{*}\leq1$. For $a^{*}>1$, there is no horizon, and the singularity is naked. The Johannsen spacetime also has the condition $a^{*}\leq1$ for the existence of an event horizon. In addition, in order to exclude pathologies such as a violation of the Lorentzian signature or the existence of closed time-like curves outside the event horizon, we impose that the metric determinant is always negative, the metric element $g_{\phi\phi}>0$ outside the event horizon, and $B$ is non-vanishing outside the horizon. These conditions lead to the following constraints on the deformation parameters $\alpha_{13}$ and $\alpha_{22}$~\citep{PhysRevD.88.044002}
\begin{align}
\alpha_{13}>&-\frac{1}{2}\left(1+\sqrt{1-a^{*2}}\right)^{4}, \label{eq:a13}
\\
-\left(1+\sqrt{1-a^{*2}}\right)^{2}&<\alpha_{22}<\frac{\left(1+\sqrt{1-a^{*2}}\right)^{4}}{a^{*2}}. \label{eq:a22}
\end{align}
%

\subsection{Accretion Disk}
\label{sec:disk}

We model the accretion disk as geometrically thin and in the equatorial plane of the BH spacetime,~i.e.~$\theta=\pi/2$ and $\dot\theta=0$, where the overhead dot represents a derivative with respect to proper time. We additionally impose that the disk is stationary and consists of particles in circular orbits. Since the spacetimes we are focusing on are stationary and axisymmetric they all possess a timelike and an azimuthal Killing vector. This in turn implies the existence of two conserved quantities: the specific energy $E$ and the $z$-component of the specific angular momentum $L_{z}$. With these conserved quantities and the imposed conditions the system is fully determined~\citep{1972ApJ...178..347B}.

The definitions of $E$ and $L_{z}$ lead to
\begin{align}
\dot t =& -\frac{Eg_{\phi\phi}+L_{z}g_{t\phi}}{g_{tt}g_{\phi\phi}-g_{t\phi}^{2}}, \label{eq:dott}
\\
\dot\phi =& \frac{Eg_{t\phi}+L_{z}g_{tt}}{g_{tt}g_{\phi\phi}-g_{t\phi}^{2}}, \label{eq:dotphi}
\end{align}
where the overhead dot represents a derivative with respect to the affine parameter (proper time for a massive particle). Substituting the above into the normalization condition for the four-velocity of massive particles $u^{a}u_{a}=-1$, we find
\begin{equation}
g_{rr}\dot r^{2}+g_{\theta\theta}\dot\theta^{2}=V_{\text{eff}}(r,\theta;E,L_{z}),
\end{equation}
where the effective potential is
\begin{equation}
V_{\text{eff}}=-1-\frac{E^{2}g_{\phi\phi}+2EL_{z}g_{t\phi}+L_{z}^{2}g_{tt}}{g_{tt}g_{\phi\phi}-g_{t\phi}^{2}},\label{eq:Veff}
\end{equation}
and the four-velocity is parametrized via $u^{a}=(\dot t, \dot r, \dot\theta, \dot\phi)$.

As we restrict our attention to equatorial and circular orbits, we can obtain explicit expressions for the energy and the angular momentum. From the stability and the circularity conditions we have $V_{\text{eff}}=0$ and $\partial V_{\text{eff}}/\partial r=0$, which allows us to solve for $E$ and $L_{z}$
\begin{align}
E=&-\left( g_{tt}+g_{t\phi}\omega \right)\dot{t}
=-\frac{g_{tt}+g_{t\phi}\omega}{\sqrt{-(g_{tt}+2g_{t\phi}\omega+g_{\phi\phi}\omega^{2})}},\label{eq:E}
\\
L_{z}=& \left( g_{t\phi}+g_{\phi\phi}\omega \right)\dot{t}
=\frac{g_{t\phi}+g_{\phi\phi}\omega}{\sqrt{-(g_{tt}+2g_{t\phi}\omega+g_{\phi\phi}\omega^{2})}},\label{eq:Lz}
\end{align}
where the angular velocity of the equatorial circular geodesics is
\begin{equation}
\omega=\frac{d\phi}{dt}=\frac{-g_{t\phi,r}\pm\sqrt{(g_{t\phi,r})^{2}-g_{tt,r}g_{\phi\phi,r}}}{g_{\phi\phi,r}},\label{eq:angvel}
\end{equation}
and
\begin{equation}
\dot t = \frac{1}{\sqrt{-(g_{tt}+2g_{t\phi}\omega+g_{\phi\phi}\omega^{2})}}.\label{eq:tdot}
\end{equation}

\begin{figure*}
\begin{center}
\includegraphics[width=8.5cm]{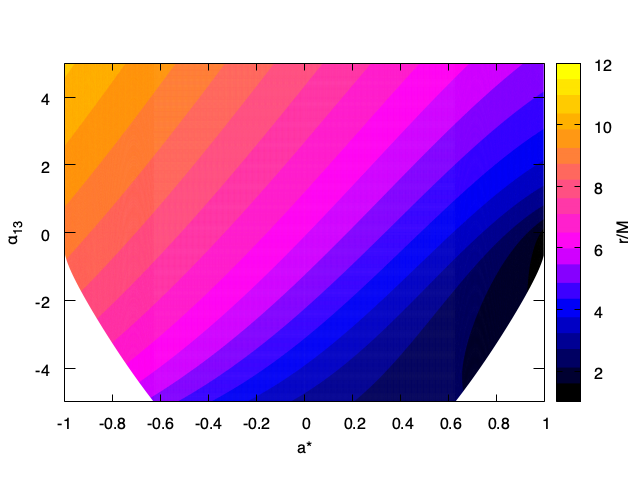}
\includegraphics[width=8.5cm]{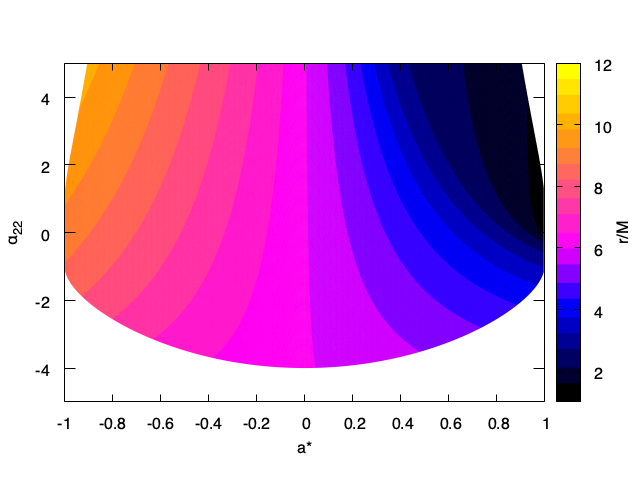}
\\
\includegraphics[width=8.5cm]{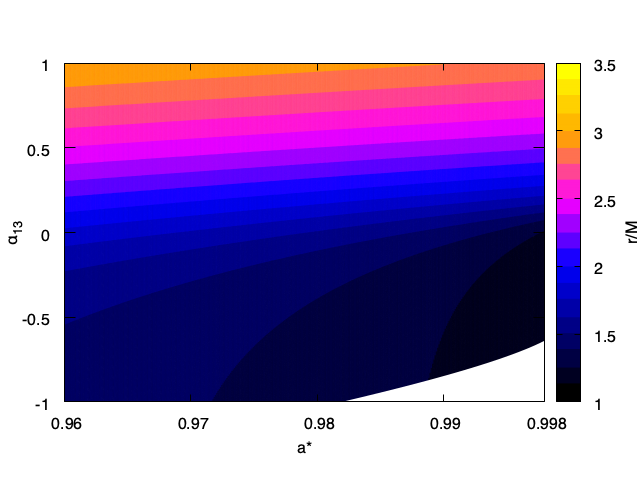}
\includegraphics[width=8.5cm]{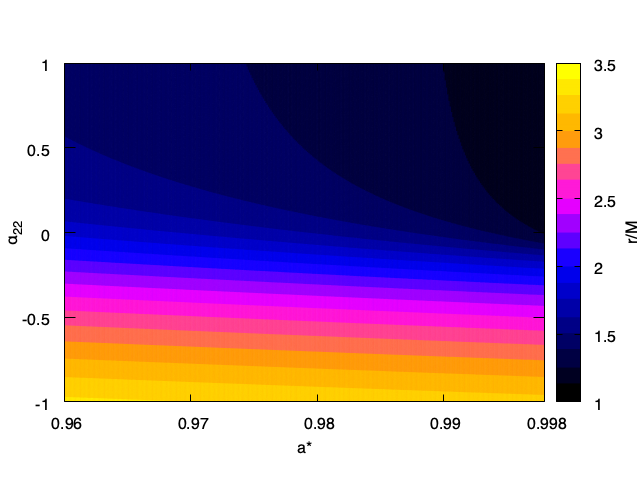}
\end{center}
\caption{\label{fig:ISCO} Contour plots of the ISCO radius for the Johannsen spacetime as a function of dimensionless spin parameter $a^{*}$ and only one non-vanishing deformation parameter $\alpha_{13}$ (left) or $\alpha_{22}$ (right). Bottom row zooms in on the high spin region near the Kerr case. Positive $a^{*}$ corresponds to a co-rotating disk and negative $a^{*}$ corresponds to a counter-rotating disk. The white regions are excluded as they violate Eqs.~\ref{eq:a13} and~\ref{eq:a22}.}
\vspace{0.3cm}
\end{figure*}

We can also calculate the innermost stable circular orbit (ISCO) of massive particles in the disk. Any circular orbit within the ISCO is unstable and, in principle, any particles there will rapidly plunge and cross the event horizon. For this reason, we will assume that the inner radius of the accretion disk cannot be smaller than the ISCO radius, $R_{\text{in}}\geq R_{\text{ISCO}}$. The ISCO radius can be found by substituting Eqs.~\ref{eq:E} and~\ref{eq:Lz} into Eq.~\ref{eq:Veff}, and then solving $\partial^{2}V_{\text{eff}}/\partial r^{2}=0$ for $r$. We plot the ISCO radius for the Johannsen metric for the cases where only $\alpha_{13}$ or $\alpha_{22}$ are non-vanishing in Figure~\ref{fig:ISCO}.

%

\subsection{Cunningham Transfer Function}
\label{sec:transfer}

Here we review the formalism of the transfer function for geometrically thin and optically thick accretion disks~\citep{1975ApJ...202..788C, 1995CoPhC..88..109S, 2010MNRAS.409.1534D}. For the reflection spectrum we are interested in the observed specific intensity $I_{o}(\nu_{o})$ at frequency $\nu_{o}$. To calculate the specific intensity we must integrate over the observing screen the local specific intensity emitted from the accretion disk $I_{\nu_{e}}(r_{e},\theta_{e})$, where $\nu_{e}$, $r_{e}$, and $\theta_{e}$ are the frequency, radius of emission, and emission angle, respectively, of emitted photons in the frame where the photons were emitted. This integration can be done by first projecting the accretion disk onto a plane perpendicular to the line of sight,~i.e.~the observer's sky~\citep{1975ApJ...202..788C}.

We place the observer at spatial infinity $(r=+\infty)$ at an inclination angle $\iota$,~i.e.~the angle between the observer's line of sight and the angular momentum of the accretion disk. On the observer's plane of the sky we use Cartesian coordinates defined as $(\alpha,\beta)$, measured along the observer's line of sight perpendicular and parallel to the rotation axis of the accretion disk when projected onto the plane, respectively. The celestial coordinates in terms of the photon momentum can then be written as
\begin{equation}
\alpha=\lim_{r\rightarrow\infty}\frac{-rp^{(\phi)}}{p^{(t)}}, \quad
\beta=\lim_{r\rightarrow\infty}\frac{rp^{(\theta)}}{p^{(t)}}, \label{eq:celcoords}
\end{equation}
where $p^{(a)}$ denotes the components of the photon's four momentum with respect to a locally non-rotating reference frame~\citep{1972ApJ...178..347B}. $p^{(a)}$ and $p^{a}$ are related through a coordinate transformation (e.g.~$p^{\phi}=p^{(\phi)}/\sin\iota$). The celestial coordinates $(\alpha,\beta)$ are related to the solid angle on the observer's sky through~\citep{1975ApJ...202..788C} $d\alpha d\beta=D^{2}d\Omega$, where $D$ is the distance between the BH and observer.

We can use Liouville's theorem~\citep{1966AnPhy..37..487L}, $I_{\nu}/\nu^{3}=\text{const}.$, to obtain the specific intensity as seen by the observer. The observed flux of an accretion disk is then given by
\begin{equation}
F_{o}(\nu_{o})=\int g^{3}I_{\nu_{e}}\left(r_{e},\theta_{e}\right)d\alpha d\beta,
\end{equation}
where the redshift factor is
\begin{equation}
g=\frac{\nu_{o}}{\nu_{e}}=\frac{(p_{a}u^{a})_{o}}{(p_{b}u^{b})_{e}}.\label{eq:redshift}
\end{equation}
Here $p_{a}$ is the canonical conjugate momentum of a photon traveling from the emitter to the observer, and $u^{a}_{o}$ and $u^{a}_{e}$ are the four velocities of the observer and emitter, respectively.

Since the spacetimes we are working with are stationary and axisymmetric the photon's conjugate momentum is given by $p_{a}=(-E^{\gamma},p_{r},p_{\theta},L_{z}^{\gamma})$. We reasonably treat the observer as static, $u_{o}^{a}=(1,0,0,0)$, and the numerator of Eq.~\ref{eq:redshift} is then $(p_{a}u^{a})_{o}=-E^{\gamma}$. We have already calculated the four velocity of the orbiting emitting material in Section~\ref{sec:disk}
\begin{equation}
u_{e}^{a}=u_{e}^{t}(1,0,0,\omega),
\end{equation}
where $u_{e}^{t}=\dot t$ given by Eq.~\ref{eq:tdot} and $\omega$ is given by Eq.~\ref{eq:angvel}. The denominator of Eq.~\ref{eq:redshift} is now $(p_{a}u^{a})_{e}=\dot t(-E^{\gamma}+\omega L_{z}^{\gamma})$, and the redshift factor is
\begin{equation}
g=\frac{\sqrt{-(g_{tt}+2g_{t\phi}\omega+g_{\phi\phi}\omega^{2})}}{1-\omega b},
\end{equation}
where $b\equiv L_{z}^{\gamma}/E^{\gamma}$.

We can also compute the emission angle $\theta_{e}$, which will be necessary if the local emission of the disk is not isotropic. The normal of the disk is given by
\begin{equation}
n^{a}=(0,0,\sqrt{g^{\theta\theta}},0)|_{r_{e},\theta_{e}=\pi/2},
\end{equation}
and therefore the emission angle is given by
\begin{equation}
\cos\theta_{e}=\frac{n^{a}p_{a}}{u_{e}^{b}p_{b}}|_{e}=g\sqrt{g^{\theta\theta}}\frac{p^{e}_{\theta}}{p^{e}_{t}}, \label{eq:cose}
\end{equation}
where $p^{e}_{a}$ is the photon conjugate momentum at the emission point in the disk.

Following \cite{1975ApJ...202..788C} we define the maximum and minimum frequency ratio $g^{*}$ at a given radius of the accretion disk
\begin{equation}
g^{*}=\frac{g-g_{\text{min}}}{g_{\text{max}}-g_{\text{min}}}\in[0,1],
\end{equation}
where $g_{\text{min}}=g_{\text{min}}(r_{e},\iota)$ and $g_{\text{max}}=g_{\text{max}}(r_{e},\iota)$ are, respectively, the minimum and maximum values of the redshift factor $g$ for photons emitted at $r_{e}$ and detected by an observer with inclination angle $\iota$. 

We can now perform a coordinate transformation from $(\alpha,\beta)$ to $(r_{e},g^{*})$, which in turn allows us to carry out the integration over the accretion disk rather than the observer's sky. This coordinate transformation is simplified through the use of the \textit{transfer function}
\begin{equation}
f(g^{*},r_{e},\iota)=\frac{1}{\pi r_{e}}g\sqrt{g^{*}(1-g^{*})}\left|\frac{\partial(\alpha,\beta)}{\partial(g^{*},r_{e})}\right|,
\end{equation}
where $\left|\partial(\alpha,\beta)/\partial(g^{*},r_{e})\right|$ is the Jacobian.

Finally, using the above equations, the observed flux of the accretion disk is given by
\begin{equation}
F_{o}(\nu_{o})=\int^{R_{\text{out}}}_{R_{\text{in}}}\int^{1}_{0}\frac{\pi r_{e}g^{2}f(g^{*},r_{e},\iota)}{\sqrt{g^{*}(1-g^{*})}}I_{e}(r_{e},\theta_{e})dg^{*}dr_{e}, \label{eq:Ione}
\end{equation}
where $R_{\text{in}}$ and $R_{\text{out}}$ are, respectively, the inner and outer radii of the disk.

In general, for given values of $r_{e}$ and $\iota$, the transfer function is a closed curve parameterized by $g^{*}$. There is only one point in the disk, and in turn in the transfer function, for which $g^{*}=0$ and one point for which $g^{*}=1$. There are two curves connecting these two points, and thus there are two branches of the transfer function, $f^{(1)}(g^{*},r_{e},\iota)$ and $f^{(2)}(g^{*},r_{e},\iota)$. Equation~\ref{eq:Ione} can be rewritten as
\begin{align}
F_{o}(\nu_{o})&=\int^{R_{\text{out}}}_{R_{\text{in}}}\int^{1}_{0}\frac{\pi r_{e}g^{2}f^{(1)}(g^{*},r_{e},\iota)}{\sqrt{g^{*}(1-g^{*})}}I_{e}(r_{e},\theta^{(1)}_{e})dg^{*}dr_{e}
\nonumber \\
&+\int^{R_{\text{out}}}_{R_{\text{in}}}\int^{1}_{0}\frac{\pi r_{e}g^{2}f^{(2)}(g^{*},r_{e},\iota)}{\sqrt{g^{*}(1-g^{*})}}I_{e}(r_{e},\theta^{(2)}_{e})dg^{*}dr_{e},
\end{align}
where $\theta^{(1)}_{e}$ and $\theta^{(2)}_{e}$ are the emission angles with relative redshift factor $g^{*}$, respectively in branches 1 and 2.

\subsection{Numerical Method}
\label{sec:num}

\begin{figure*}
\begin{center}
\includegraphics[width=8.5cm]{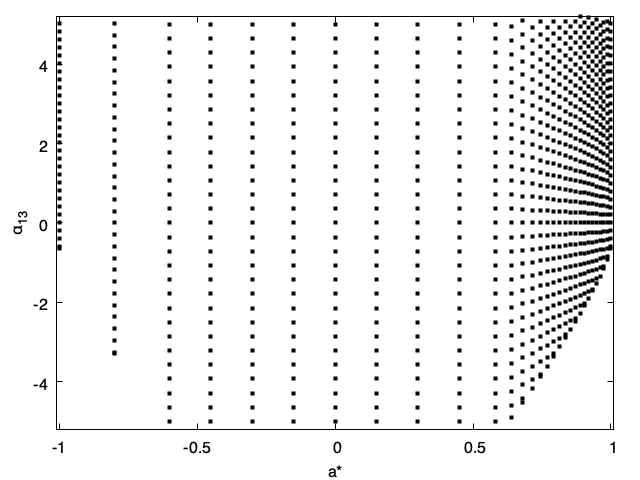}
\includegraphics[width=8.5cm]{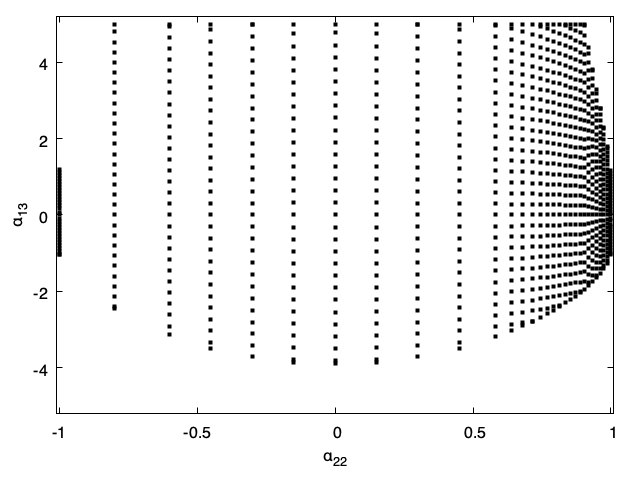}
\end{center}
\caption{\label{fig:grid} Grid points in the FITS file for dimensionless spin parameter $a^{*}$ and deformation parameters $\alpha_{13}$ (left) and $\alpha_{22}$ (right).}
\vspace{0.3cm}
\end{figure*}

Following the methodology of \textsc{relxill} we generate a FITS (Flexible Image Transport System) file containing the relevant spacetime information. The three physical parameters describing the BH spacetime in the table are the dimensionless BH spin parameter, the deformation parameter, and the inclination angle, in a grid of 30 by 30 by 22, respectively. The grid points for the BH spin are more dense towards higher spin (and prograde disk rotation) as the ISCO radius changes more rapidly as spin increases. For the deformation parameters $\alpha_{13}$ and $\alpha_{22}$ of the Johannsen metric the grid points are uniformly distributed in the range $[-5,5]$. For values of spin where the constraints on the deformation parameters given by Eqs.~\ref{eq:a13} and~\ref{eq:a22} fall into this range, the range is adjusted to obey the constraints. Figure~\ref{fig:grid} shows the distribution of grid points in the spin-deformation parameter phase space{\footnote{When \textsc{relxill\_nk} is used within \textsc{xspec} the deformation parameter values are scaled to be in the range $[-1,1]$ for each value of spin. The values must be unscaled outside of \textsc{xspec}. This is done because the constraints on the deformation parameters in the Johannsen metric in Eqs.~\ref{eq:a13} and~\ref{eq:a22} (similar behavior is possible in other metrics) lead to a spin-dependent allowed range for the deformation parameters. It is difficult to incorporate such a range directly into \textsc{xspec}.}}. The grid points for the inclination angle are distributed evenly in $0<\cos\iota<1$. For each set of physical parameters, the accretion disk is discretized into a grid of 100 emission radii $r_{e}$ and for each $r_{e}$ the transfer function is tabulated at 20 equally spaced values of $g^{*}$ on each branch of the transfer function. The emission angle is also calculated and tabulated for each of these accretion disk grid points.

We use a general relativistic ray-tracing code to calculate the Jacobian, redshift factor, and emission angle, necessary for the FITS file. Our ray-tracing code computes the trajectories of photons from the BH accretion disk to a distant observer following the method described in~\cite{Psaltis:2010ww} and is a modified version of the code used in~\cite{Ayzenberg:2018jip, Gott:2018ocn}. As explained previously, all stationary and axisymmetric spacetimes have conserved energy $E$ and angular momentum $L_{z}$ that are related to the four-momentum of a test particle: $p_{t}=-E$ and $p_{\phi}=L_{z}$. This leads to two first-order differential equations shown in Eqs.~\ref{eq:dott} and~\ref{eq:dotphi}, which we rewrite as
\begin{align}
\frac{dt}{d\lambda'} =& -\frac{bg_{t\phi} + g_{\phi\phi}}{g_{tt}g_{\phi\phi}-g_{t\phi}^{2}}, \label{eq:dt}
\\
\frac{d\phi}{d\lambda'} =& b\frac{g_{t\phi}+g_{tt}}{g_{tt}g_{\phi\phi}-g_{t\phi}^{2}}, \label{eq:dp}
\end{align}
where $\lambda'\equiv E/\lambda$ is the normalized affine parameter and $b\equiv L_{z}/E$ is the impact parameter.

The $r-$ and $\theta-$components of the photon position are described through the second-order geodesic equations for a generic axisymmetric metric
\begin{widetext}
\begin{align}
\frac{d^{2}r}{d\lambda'^{2}}=&-\Gamma^{r}_{tt}\left(\frac{dt}{d\lambda'}\right)^{2}-\Gamma^{r}_{rr}\left(\frac{dr}{d\lambda'}\right)^{2}-\Gamma^{r}_{\theta\theta}\left(\frac{d\theta}{d\lambda'}\right)^{2}-\Gamma^{r}_{\phi\phi}\left(\frac{d\phi}{d\lambda'}\right)^{2}-2\Gamma^{r}{t\phi}\left(\frac{dt}{d\lambda'}\right)\left(\frac{d\phi}{d\lambda'}\right)-2\Gamma^{r}_{r\theta}\left(\frac{dr}{d\lambda'}\right)\left(\frac{d\theta}{d\lambda'}\right), \label{eq:d2r}
\\
\frac{d^{2}\theta}{d\lambda'^{2}}=&-\Gamma^{\theta}_{tt}\left(\frac{dt}{d\lambda'}\right)^{2}-\Gamma^{\theta}_{rr}\left(\frac{dr}{d\lambda'}\right)^{2}-\Gamma^{\theta}_{\theta\theta}\left(\frac{d\theta}{d\lambda'}\right)^{2}-\Gamma^{\theta}_{\phi\phi}\left(\frac{d\phi}{d\lambda'}\right)^{2}-2\Gamma^{\theta}{t\phi}\left(\frac{dt}{d\lambda'}\right)\left(\frac{d\phi}{d\lambda'}\right)-2\Gamma^{\theta}_{r\theta}\left(\frac{dr}{d\lambda'}\right)\left(\frac{d\theta}{d\lambda'}\right), \label{eq:d2th}
\end{align}
\end{widetext}
where $\Gamma^{a}_{bc}$ are the Christoffel symbols of the metric.

We choose a coordinate system and reference frame such that the BH is stationary at the origin and the BH's spin angular momentum is along the $z$-axis. As the reflection spectrum is independent of the BH mass $M$, in this code and for the remainder of this paper, we use units with $M=1$. For the numerical evolution, the observing screen is centered at a distance $D=10^{8}$, the polar angle $\theta=\iota$, and the azimuthal angle $\phi=0$. On the screen, we use polar coordinates $r_{\text{scr}}$ and $\phi_{\text{scr}}$, which relate to the celestial coordinates of Eq.~\ref{eq:celcoords} via $\alpha=r_{\text{scr}}\cos\phi_{\text{scr}}$ and $\beta=r_{\text{scr}}\sin\phi_{\text{scr}}$.

We solve the system of equations (Eqs.~\ref{eq:dt}-\ref{eq:d2th}) backwards in time, initializing each photon with an initial position and a four-momentum that is perpendicular to the screen. The latter simulates placing the observing screen at spatial infinity as only photons traveling perpendicular to the screen at distance $D$ will also impact the screen at spatial infinity.

The initial position and four-momentum of each photon in the BL coordinates of the BH spacetime is given by
\begin{align}
r_{i}=&\left(\alpha^{2}+\beta^{2}+D^{2}\right)^{1/2},
\\
\theta_{i}=&\arccos\left(\frac{D\cos\iota+\beta\sin\iota}{r_{i}}\right),
\\
\phi_{i}=&\arctan\left(\frac{\alpha}{D\sin\iota-\beta\cos\iota}\right),
\end{align}
and
\begin{align}
\left(\frac{dr}{d\lambda'}\right)_{i}=&\frac{D}{r_{i}},
\\
\left(\frac{d\theta}{d\lambda'}\right)_{i}=&\frac{-\cos\iota+\frac{d}{r_{i}^{2}}\left(D\cos\iota+\beta\sin\iota\right)}{\sqrt{r_{i}^{2}-\left(D\cos\iota+\beta\sin\iota\right)^{2}}},
\\
\left(\frac{d\phi}{d\lambda'}\right)_{i}=&\frac{-\alpha\sin\iota}{\alpha^{2}+\left(D\sin\iota-\beta\cos\iota\right)^{2}},
\\
\left(\frac{dt}{d\lambda'}\right)_{i}=&\frac{g_{t\phi}}{g_{tt}}\left(\frac{d\phi}{d\lambda'}\right)_{i}-\left[\frac{g_{t\phi}^{2}}{g_{tt}^{2}}\left(\frac{d\phi}{d\lambda'}\right)_{i}^{2}-\left(g_{rr}\left(\frac{dr}{d\lambda'}\right)_{i}^{2}\right.\right.
\nonumber \\
&\left.\left.+g_{\theta\theta}\left(\frac{d\theta}{d\lambda'}\right)_{i}^{2}+g_{\phi\phi}\left(\frac{d\phi}{d\lambda'}\right)_{i}^{2}\right)\right]^{1/2}.
\end{align}
Requiring that the norm of the photon four-momentum is zero provides the last component $\left(dt/d\lambda'\right)_{i}$. As the impact parameter $b$ is a conserved quantity and is required in Eqs.~\ref{eq:dt} and~\ref{eq:dp}, it is computed from the initial conditions.

We use an adaptive algorithm to search for the photons that hit the accretion disk,~i.e.~the $\theta=\pi/2$ plane, at the 100 disk emission radii $r_{e}$ to within a precision of $\sim10^{-6}$ by varying $r_{\text{scr}}$. For each emission radius we find at least 62 different photons by varying $\phi_{\text{scr}}$ in equally spaced values in the range $[0,2\pi]$. Two additional adaptive algorithms are used to find $g_{\text{min}}$ and $g_{\text{max}}$ and then to better fill the $g^{*}$ space if necessary.

For each of these photons the redshift factor $g$ (Eq.~\ref{eq:redshift}), emission angle $\theta_{e}$ (Eq.~\ref{eq:cose}), and Jacobian $|\partial(\alpha,\beta)/\partial(g^{*},r_{e})|$ are calculated. To calculate the latter we use
\begin{equation}
\left|\frac{\partial(\alpha,\beta)}{\partial(g^{*},r_{e})}\right|=\left(g_{\text{max}}-g_{\text{min}}\right)\left|\frac{\partial\alpha}{\partial g}\frac{\partial\beta}{\partial r_{e}}-\frac{\partial\alpha}{\partial r_{e}}\frac{\partial\beta}{\partial g}\right|, \label{eq:jac}
\end{equation}
where the first term on the right-hand side is computed in a separate code afterwards and the second term is computed by solving the geodesic equations for an additional four photons. These four photons are initialized on the screen at $(\alpha_{0}\pm\Delta\alpha,\beta_{0}\pm\Delta\beta)$, where $(\alpha_{0},\beta_{0})$ are the initial coordinates of the original photon, $\Delta\alpha=10^{-5}+10^{-5}\alpha_{0}$, and $\Delta\beta=10^{-5}+10^{-5}\beta_{0}$. The derivatives in are then approximated from the emission radius, redshift factor, and initial coordinates of these four photons.

The adaptive algorithm to find $g_{\text{min}}$ and $g_{\text{max}}$ starts from the initial 62 photons for a given $r_{e}$, from which we record preliminary $g_{\text{min}}$ and $g_{\text{max}}$. Using an adaptive step-size we shift $\phi_{\text{scr}}$ from these preliminary redshift extrema and search for the actual extrema. Once the change in the redshift factor between consecutive steps is less than $10^{-6}$, we stop the algorithm and set this photon and its related redshift factor as the extrema.

The adaptive algorithm to better fill the $g^{*}$ space calculates $g^{*}$ for every photon and compares the values between consecutive photons. If the difference between consecutive $g^{*}$'s is greater than 0.05, a search for an additional photon with $g^{*}$ between the two is performed.

Finally, a separate script is used to process all photons and create the FITS file. The data is split into two branches according to
\begin{equation}
\phi_{\text{scr}}^{\text{min}}<\phi_{\text{scr}}<\phi_{\text{scr}}^{\text{max}} \quad
\text{and} \quad
\phi_{\text{scr}}^{\text{min}}>\phi_{\text{scr}}>\phi_{\text{scr}}^{\text{max}}
\end{equation}
where $\phi_{\text{scr}}^{\text{min}}$ and $\phi_{\text{scr}}^{\text{max}}$ correspond to the photons for $g_{\text{min}}$ and $g_{\text{max}}$, respectively. Then, a linear interpolation is used to calculate 20 values of the transfer function at equally spaced values of $g^{*}$ for each branch. The emission angles $\theta_{e}$ at each $g^{*}$ are also computed using a linear interpolation. A FITS file containing the values of emission radius $r_{e}$, extrema redshift $g_{\text{min}}$ and $g_{\text{max}}$, transfer functions, and emission angles $\theta_{e}$, for the full set of physical parameters dimensionless spin $a^{*}$, deformation parameter, and inclination angle $\iota$, is generated at the end.

\subsection{Lamppost Geometry}
\label{sec:lamp}

The base versions of \textsc{relxill} and \textsc{relxill\_nk} make no strict assumptions about the geometry and location of the hot corona and instead assume the impinging radiation on the disk is a power-law or broken power-law. An alternative model implemented in \textsc{relxilllp}~\citep{doi:10.1093/mnras/sts710} treats the corona as a isotropically-emitting point source at height $h$ along the spin axis of the BH. The impinging radiation profile on the disk is determined by solving for the photon trajectory in the spacetime. As with the transfer function calculation, within the Kerr spacetime that \textsc{relxill} assumes, the calculation of the impinging radiation profile for \textsc{relxilllp} reduces to numerically integrating two elliptical integrals. For the non-Kerr version, \textsc{relxilllp\_nk}, we use the general relativistic ray-tracing code described in Sec.~\ref{sec:num} to calculate the relevant quantities.

For the lamppost geometry we create an additional FITS file to store the necessary information about the impinging radiation. This file has a similar structure as that of the Master Table FITS file described in Sec.~\ref{sec:num}, but the inclination angle is replaced by the height and the stored data consists of the incident intensity $I_{i}$, the angle of emission from the corona $\delta$, and the incident angle $\delta_{i}$, for 100 values of emission radius $r_{e}$. The height varies from the vicinity of the horizon radius up to 500 in a grid of 250 values.

In order to calculate the incident intensity $I_{i}$, we use ray-tracing to calculate the trajectories of 12,000 photons emitted from the corona point-source with equally spaced emission angles $\delta$. Each trajectory is stopped at the accretion disk in the $\theta=\pi/2$ plane, providing an incident location $(r_{i}, \delta_{i})$ for each photon. With the incident location for each photon we can calculate the photon flux incident on the accretion disk. Since the photons are emitted isotropically in equally spaced angles, the distance $\Delta r_{i}$ between incident locations is related to the incident intensity. Photons emitted in the range $[\delta,\delta+\Delta\delta]$ impact the disk in a ring with area $A(r,\Delta r)$. The proper area of such a ring is
\begin{equation}
A(r,\Delta r)=2\pi\sqrt{g_{rr}g_{\phi\phi}}\Delta r,
\end{equation}
in the rest frame of the observer~\citep{2012MNRAS.424.1284W}.

In the rest frame of the accretion disk, we must include the effect of the disk's rotation. The Lorentz factor of the disk is given by~\citep{1972ApJ...178..347B}
\begin{equation}
\gamma=\left[\frac{\left(\omega-\frac{g_{t\phi}}{g_{\phi\phi}}\right)^{2}g_{\phi\phi}^{2}}{g_{tt}g_{\phi\phi}-g_{t\phi}^{2}}+1\right]^{-1/2},
\end{equation}
where $\omega$ is the disk's angular velocity given by Eq.~\ref{eq:angvel}.

Factoring in that the emission is isotropic, the incident intensity is then
\begin{equation}
I_{i}=\frac{\sin\delta}{A(r,\Delta r)\gamma}.
\end{equation}
Due to the relative motion of the corona and the accretion disk, as well as the general relativistic effects, the incident spectrum will be shifted in energy relative to the emitted spectrum~\citep{2007ApJ...664...14F}. The redshift factor here is calculated in the same way as that in Eq.~\ref{eq:redshift}, however the four velocities of the emitting material and the observer are reversed,~i.e.~the corona is static, $u^{a}_{c}=(1,0,0,0)$, and the observer is the rotating disk, $u^{a}_{d}=u^{t}_{d}(1,0,0,\omega)$. The lamppost redshift factor is then given by
\begin{equation}
g_{lp}=\frac{E_{i}}{E_{e}}=\frac{p_a u^{a}_{d}}{p_{b}u^{b}_{c}}=\sqrt{\frac{g_{tt}|_{c}}{\left(g_{tt}+2g_{t\phi}\omega+g_{\phi\phi}\omega^{2}\right)|_{d}}},
\end{equation}
where the numerator within the radical is evaluated at the corona and the denominator is evaluated at the incident location on the disk.

Assuming a power-law for the emitted radiation from the corona, the incident flux on the disk is
\begin{equation}
F_{i}(r,h)=I_{i}g_{lp}^{\Gamma}=\frac{\sin\delta g_{lp}^{\Gamma}}{A(r,\Delta r)\gamma},\label{eq:Finc}
\end{equation}
where $\Gamma$ is the power law index. This incident flux is what replaces the power-law flux that is used in the \textsc{relxill} and \textsc{relxill\_nk} models.

The incident angle $\delta_{i}$ is also important as it determines the interaction depth of the reflected photon that is incorporated by the $\textsc{xillver}$ part of the model. The incident angle is found the same way as the emission angle $\theta_{e}$ in Eq.~\ref{eq:cose} and is given by
\begin{equation}
\cos\delta_{i}=\frac{n^{a}p_{a}}{u^{b}_{d}p_{b}}|_{d}=g\sqrt{g^{\theta\theta}}\frac{p^{d}_{\theta}}{p^{d}_{t}},
\end{equation}
where the emitting material in the disk is now the absorbing material in the disk.

\section{Comparison to \textsc{relxill}}
\label{sec:comp}

Here we compare test spectra produced by \textsc{relxill} to those produced by \textsc{relxill\_nk} in the Kerr spacetime by setting $\alpha_{13}=\alpha_{22}=0$ to show the accuracy of the ray-tracing method used in the latter. We only compare \textsc{relxill}/\textsc{relxill\_nk} and \textsc{relxilllp}/\textsc{relxilllp\_nk}, as the \textsc{relline} models are only for a single line, while we are interested primarily in the full reflection spectrum, and the other models available do not further modify the gravitational physics in which we are interested.

\begin{figure*}
\begin{center}
\includegraphics[width=17.0cm]{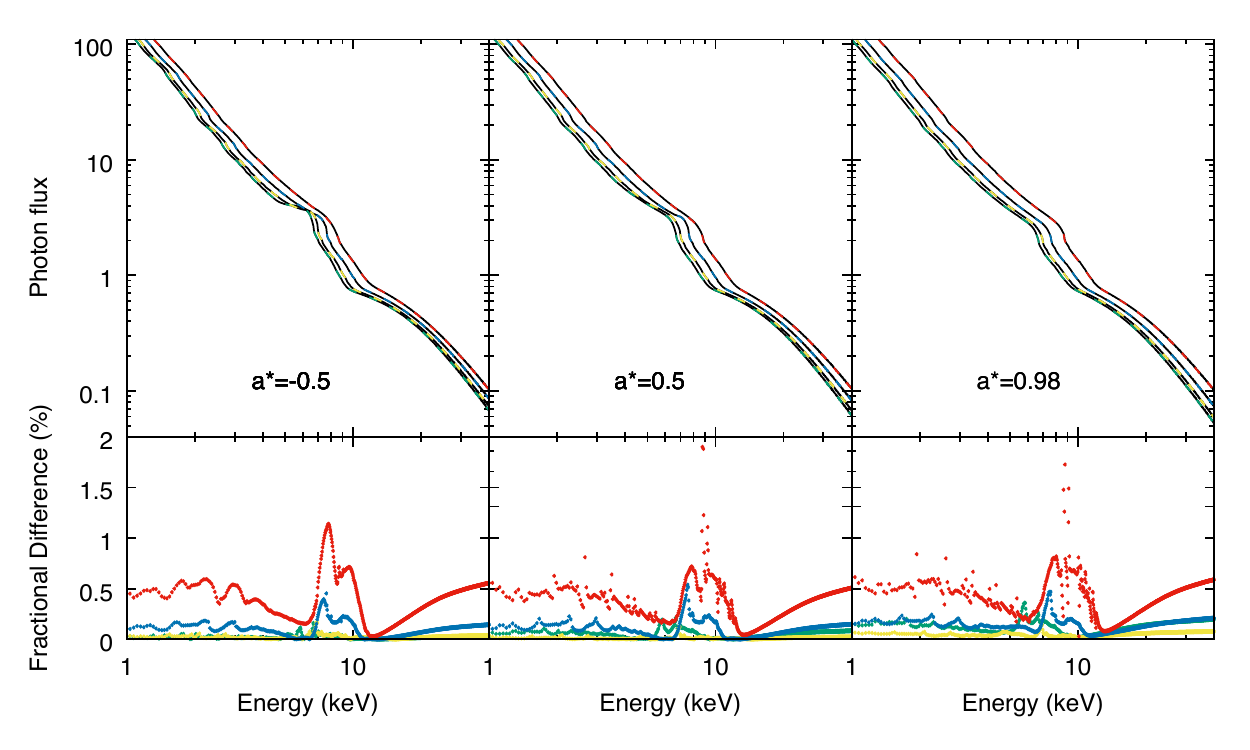}
\end{center}
\caption{\label{fig:compxill} Comparison of \textsc{relxill} (black, solid lines) and \textsc{relxill\_nk} (colored, dashed lines) for the Kerr spacetime for dimensionless spin $a^{*}=[-0.5$(left), $0.5$(center), $0.98$(right)] and inclination angle $\iota=[10\degree$(green), $30\degree$(yellow), $50\degree$(blue), $70\degree$(red)]. Other model parameters are shown in Table~\ref{tab:params}.}
\end{figure*}

\begin{figure*}
\begin{center}
\includegraphics[width=17.0cm]{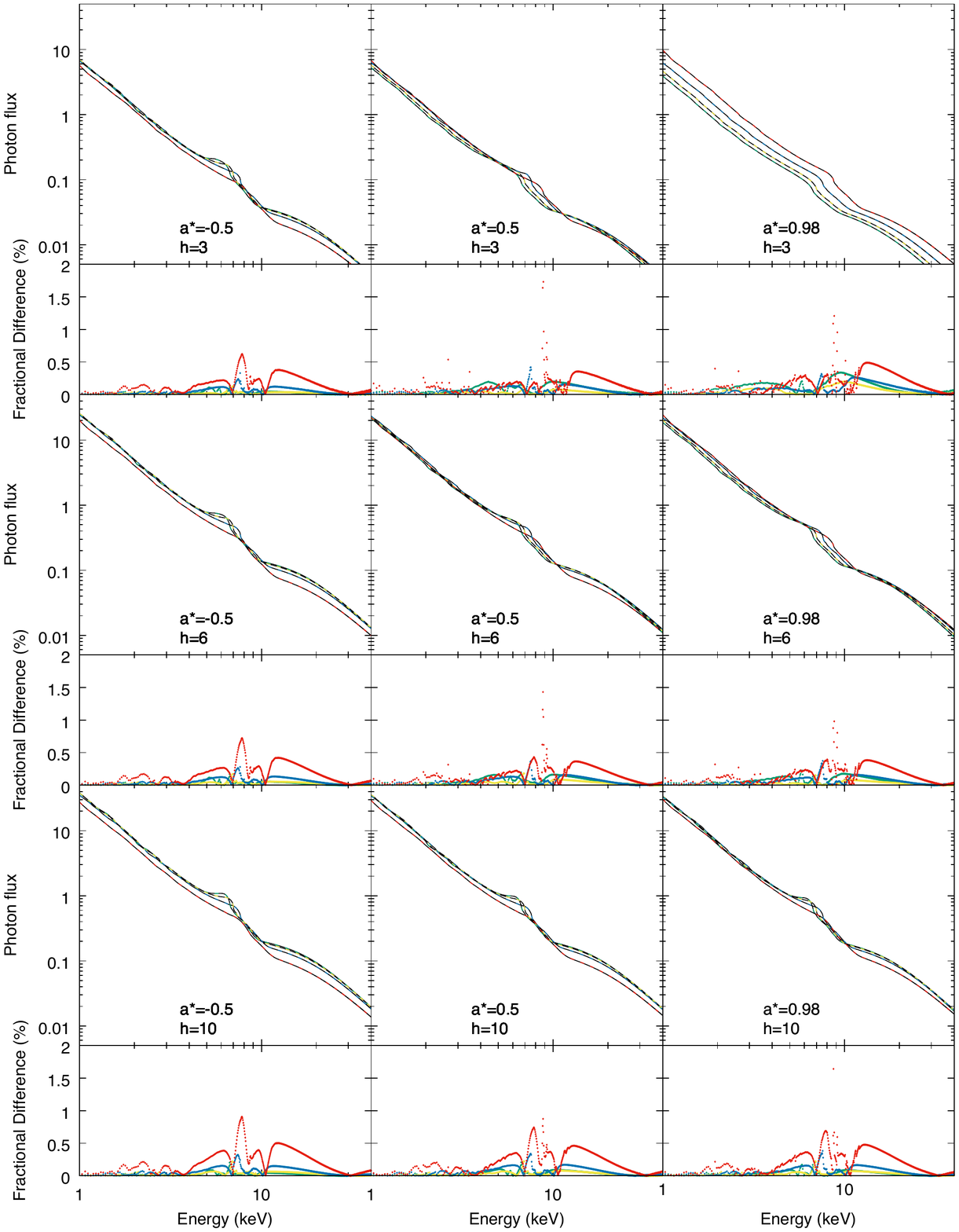}
\end{center}
\caption{\label{fig:compxilllp} Comparison of \textsc{relxilllp} (black, solid lines) and \textsc{relxilllp\_nk} (colored, dashed lines) for the Kerr spacetime for dimensionless spin $a^{*}=[-0.5$(left), $0.5$(center), $0.98$(right)], inclination angle $\iota=[10\degree$(green), $30\degree$(yellow), $50\degree$(blue), $70\degree$(red)], and height $h=[3$(top), $6$(middle), $10$(bottom)]. Other model parameters are shown in Table~\ref{tab:params}.}
\end{figure*}

We generate the spectra using \textsc{xspec} v.12.9.1p with \textsc{relxill} v.1.2.0 and \textsc{relxill\_nk} v.1.3.2. We compare spectra for dimensionless spin $a^{*}=[-0.5, 0.5, 0.98]$ and inclination angle $\iota=[10\degree, 30\degree, 50\degree, 70\degree]$. For the lamppost corona models we use height $h=[3,6,10]$. The other model parameters are kept the same (see Table~\ref{tab:params}). We calculate the fractional difference between the Kerr and non-Kerr models (fractional difference is given by $|L_{\text{K}}(\nu)-L_{\text{NK}}(\nu)|/L_{\text{K}}(\nu)$, where $L_{\text{K}}$ and $L_{\text{NK}}$ are the Kerr and non-Kerr luminosities, respectively) to show the accuracy of our new set of models, assuming the Kerr model is more accurate as the calculation is overall simpler. The resulting spectra and fractional differences are shown in Figs.~\ref{fig:compxill} and~\ref{fig:compxilllp}{\footnote{Note that \textsc{relxilllp} has a different normalization than \textsc{relxilllp\_nk}. We have renormalized the \textsc{relxilllp\_nk} reflection spectra in Fig.~\ref{fig:compxilllp} such that it matches that of the \textsc{relxilllp} spectra.}}.

Our new non-Kerr models match the Kerr models fairly well. We find that the fractional difference is at most $2\%$, but is usually below $1\%${\footnote{The accuracy in \textsc{relxill\_nk} and related models becomes significantly poor at inclination angles $\iota\gtrsim75\degree$ in the currently available FITS files (v1.2)}}. As current observational data of BH reflection spectra leads to spin estimates with errors of roughly $10\%$~\citep{2017RvMP...89b5001B} and that it is likely systematic errors in the modeling are significantly larger (see e.g.~\cite{Taylor:2017jep} and~\cite{Kammoun:2018ltv}), it is fair to say that the numerical error present in \textsc{relxill\_nk} and related models is small enough for the purposes of analyzing observational data with the new models presented in this work. We speculate that the primary sources of error are the calculation of the Jacobian shown in Eq.~\ref{eq:jac} and the error of the geodesic integrator itself. However, it is not straightforward to determine the precise sources of error or their impact as the output of the ray-tracing code goes through several layers of interpolation and integration to produce the spectrum.

\begin{table*}
\centering
\vspace{0.5cm}
\begin{tabular}{l|ccccccccccccc}
\hline\hline

& $q_{\rm in}$ & $q_{\rm out}$ & $R_{\rm br}$ & $R_{\rm in}$ & $R_{\rm out}$ & $z$ & $\Gamma$ & $\log\xi$ & $A_{\rm Fe}$ & $E_{\rm cut}$ & $R_{\rm f}$\\

\hline

{\sc relxill}/{\sc relxill\_nk} (Figs.~\ref{fig:compxill} \&~\ref{fig:relxill}) & 3 & 3 & 15 & $-1$ & 400 & 0 & 2 & 3.1 & 1 & 300 & $-1$ \\ 

{\sc relxilllp}/{\sc relxilllp\_nk} (Figs.~\ref{fig:compxilllp} \&~\ref{fig:relxilllp}) & -- & -- & -- & $-1$ & 400 & 0 & 2 & 3.1 & 1 & 300 & $-1$ \\ 

\hline\hline
\end{tabular}
\vspace{0.2cm}
\caption{\label{tab:params} Model parameters used for Figs.~\ref{fig:compxill}-\ref{fig:relxilllp}. Model parameters not shown here are stated in the captions of the figures. $R_{in}=-1$ corresponds to setting $R_{in}$ to the ISCO radius and $R_f=-1$ corresponds to only including the reflected component of the spectrum.}
\end{table*}
\vspace{0.5cm}
%

\section{Reflection Line/Spectrum Models}
\label{sec:mods}

Here we briefly describe the different models available in the \textsc{relxill\_nk} package and how introducing a non-Kerr modification to the spacetime modifies the observed spectrum. Table~\ref{tab:flavors} lists all of the available models and the parameters of each model. In the following we briefly summarize each model:

\begin{itemize}
\item \textsc{relline\_nk}: Base non-Kerr version of relativistic line model \textsc{relline}.
\item \textsc{relconv\_nk}: Similar to \textsc{relline\_nk}, but can convolve any reflection.
\item \textsc{relxill\_nk}: Base non-Kerr version of relativistic reflection model \textsc{relxill}, in which the irradiation of the disk is modeled by a broken power-law emissivity.
\item \textsc{relxillCp\_nk}: Modification of \textsc{relxill\_nk} that uses an nthcomp Comptonization~\citep{1996MNRAS.283..193Z, 1999MNRAS.309..561Z} continuum for the incident spectrum.
\item \textsc{relxillD\_nk}: Same as \textsc{relxill\_nk}, but allows for higher accretion disk electron density (between $10^{15}$ and $10^{19}$ cm$^{-3}$) and the energy cutoff $E_{\text{cut}}=300$ keV.
\item \textsc{rellinelp\_nk}: Modification of \textsc{relline\_nk} in which the incident flux on the disk is due to a isotropically emitting point source at some height along the spin axis of the BH.
\item \textsc{relxilllp\_nk}: Modification of \textsc{relxill\_nk} in which the incident flux on the disk is due to a isotropically emitting point source at some height along the spin axis of the BH.
\item \textsc{relxilllpCp\_nk}: Modification of \textsc{relxilllp\_nk} in which the incident spectrum is an nthcomp Comptonization continuum.
\item \textsc{relxilllpD\_nk}: Same as \textsc{relxilllp\_nk}, but allows for a higher accretion disk electron density (between $10^{15}$ and $10^{19}$ cm$^{-3}$) and the energy cutoff $E_{\text{cut}}=300$ keV.
\end{itemize}

\begin{figure*}
\begin{center}
\includegraphics[width=17.0cm]{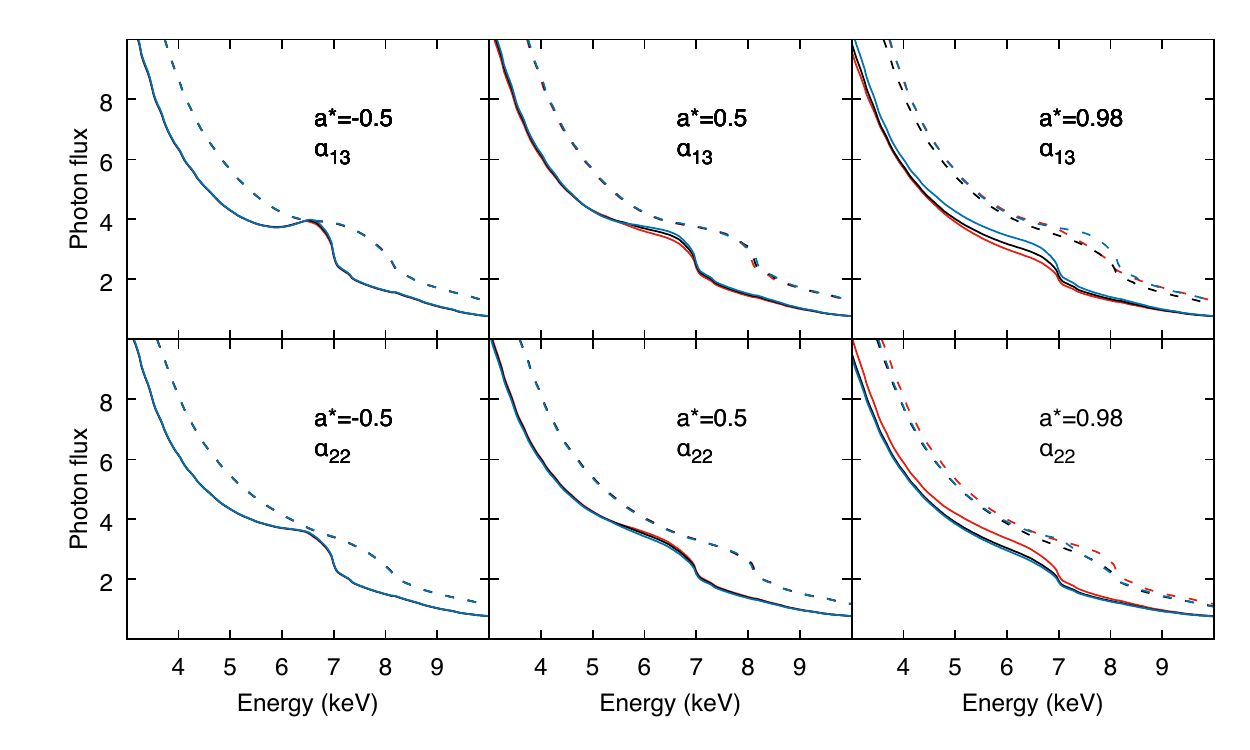}
\end{center}
\vspace{-0.3cm}
\caption{\label{fig:relxill} Comparison of \textsc{relxill\_nk} in the Johannsen spacetime for dimensionless spin $a^{*}=[-0.5$(left), $0.5$(center), $0.98$(right)] and inclination angle $\iota=[30\degree$(solid), $60\degree$(dashed)]. The top row has $\alpha_{13}=[-1$(red), $0$(black), $1$(blue)] and the bottom row has $\alpha_{22}=[-1$(red), $0$(black), $1$(blue)]. Other model parameters are shown in Table~\ref{tab:params}.}
\end{figure*}

\begin{figure*}
\begin{center}
\includegraphics[width=17.0cm]{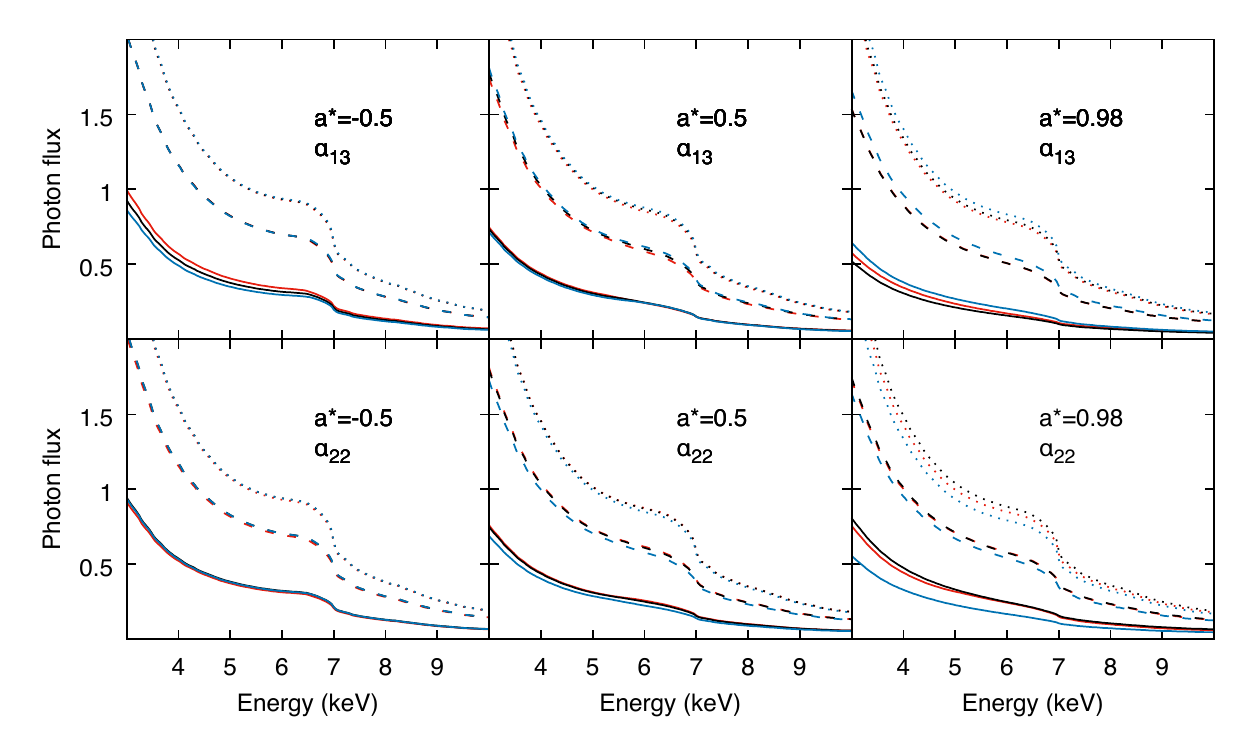}
\end{center}
\vspace{-0.3cm}
\caption{\label{fig:relxilllp} Comparison of \textsc{relxilllp\_nk} in the Johannsen spacetime for dimensionless spin $a^{*}=[-0.5$(left), $0.5$(center), $0.98$(right)], inclination angle $\iota=30\degree$, and height $h=[3$(solid), $6$(dashed), $10$(dotted)]. The top row has $\alpha_{13}=[-1$(red), $0$(black), $1$(blue)] and the bottom row has $\alpha_{22}=[-1$(red), $0$(black), $1$(blue)]. Other model parameters are shown in Table~\ref{tab:params}.}
\end{figure*}

We compare spectra in the Johannsen spacetime using the \textsc{relxill\_nk} and \textsc{relxilllp\_nk} models in Figs.~\ref{fig:relxill} and~\ref{fig:relxilllp}. For all models we use dimensionless spin $a^{*}=[-0.5,0.5,0.98]$, deformation parameters $\alpha_{13}=[-1,0,1]$ or $\alpha_{22}=[-1,0,1]$, $\iota=30\degree$, and height $h=[3,6,10]$. The other model parameters are given in Table~\ref{tab:params}. Note that we have zoomed in on the region where the K$\alpha$ line is present as this is where the non-Kerr modifications are most apparent.

It is clear from the spectra that higher values of spin increase the effect of the non-Kerr modifications, i.e.~the shape of the K$\alpha$ line region is more significantly modified by the non-Kerr deformation parameters as spin increases. For spins of $a^{*}=-0.5$ the modification is barely visible, while there is a clear difference in the spectra for spins of $a^{*}=0.98$. This is likely primarily due to the ISCO radius being smaller for higher values of spin, which in turn accentuates the strong gravity non-Kerr modifications. At smaller values of spin (and retrograde accretion disks) the ISCO radius is larger and the non-Kerr modifications are less noticeable.

\begin{figure*}
\begin{center}
\includegraphics[width=17.0cm]{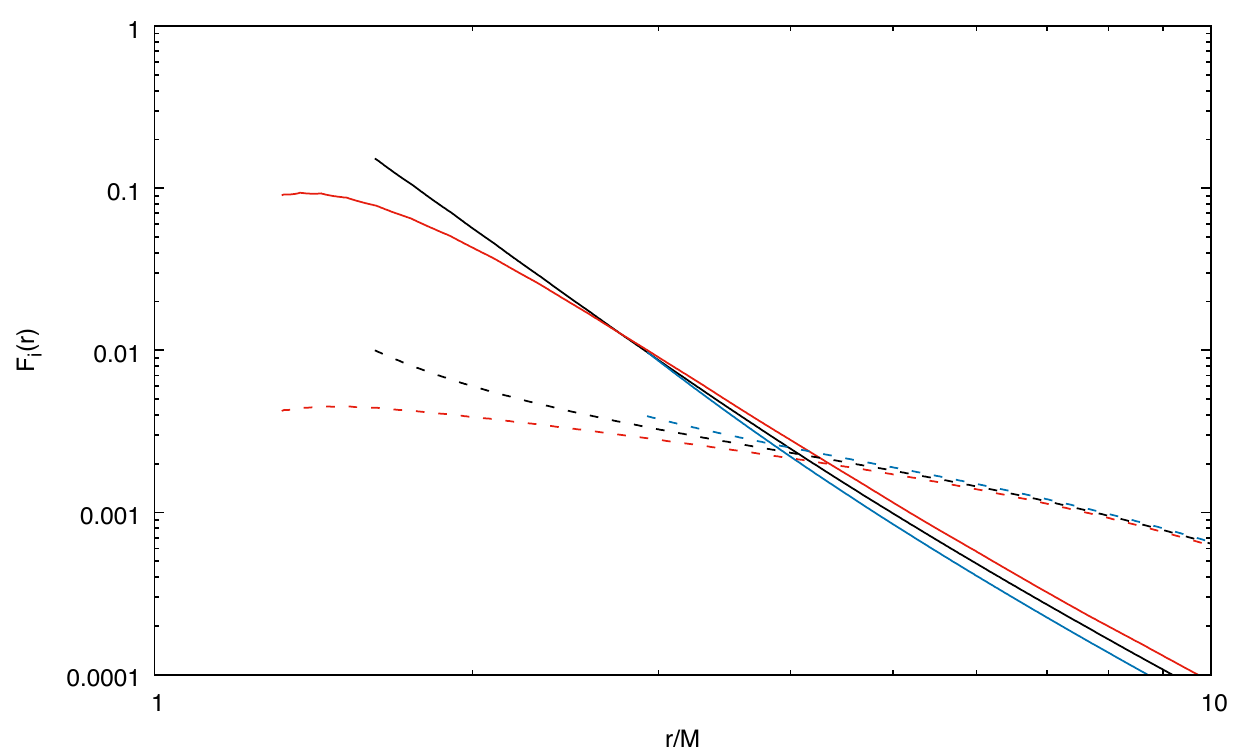}
\end{center}
\caption{\label{fig:flux} Comparison of the incident flux on the disk $F_{i}(r)$ (Eq.~\ref{eq:Finc}) in the Johannsen spacetime for dimensionless spin $a^{*}=0.98$, inclination angle $\iota=30\degree$, lamppost height $h=[2\text{(solid)},10\text{(dashed)}]$, and non-zero deformation parameter $\alpha_{13}=[-1$(red), $0$(black), $1$(blue)]. Other model parameters are shown in Table~\ref{tab:params}.}
\end{figure*}

In the lamppost corona model \textsc{relxilllp\_nk}, one would naively expect the non-Kerr modifications to be more significant at smaller values of height as more of the photons emitted by the corona must travel through the strong gravity region very near the BH. However, this seems to not be the case, as the modifications at different values of height are of roughly equivalent magnitude (we have checked this for values of height down to $h=2$). The explanation for this can be seen in Fig.~\ref{fig:flux} where we plot the incident flux on the disk $F_{i}(r)$ given by Eq.~\ref{eq:Finc} for two values of height $h=[2,10]$ and three values of deformation parameter $\alpha_{13}=[-1,0,1]$. Notice that at both values of height the incident flux in the non-Kerr cases only shows significant departure from the Kerr case for very small radius $r\lesssim2$ and the magnitude of the departure is comparable at both values of height. Thus, any non-Kerr modifications to the spectrum due to the modifications in the incident flux seem to be lamppost height independent and are suppressed by the lack of significant modifications over disk radii larger than $r\approx2$.

\begin{table*}
\centering
\vspace{0.5cm}
\begin{tabular}{l|cccccccccccccccccccc}
\hline\hline

& \hspace{-0.25cm} $E_{\rm line}$  \hspace{-0.25cm} &  \hspace{-0.25cm} $q_{\rm in}$  \hspace{-0.25cm}&  \hspace{-0.25cm} $q_{\rm out}$  \hspace{-0.25cm} &  \hspace{-0.25cm} $R_{\rm br}$  \hspace{-0.25cm} &  \hspace{-0.25cm}$h$  \hspace{-0.25cm} &  \hspace{-0.25cm} $a^*$ \hspace{-0.25cm} & \hspace{-0.25cm} $\iota$ \hspace{-0.25cm} & \hspace{-0.25cm} $R_{\rm in}$ \hspace{-0.25cm} & \hspace{-0.25cm} $R_{\rm out} \hspace{-0.25cm}$ & \hspace{-0.25cm} $z$ \hspace{-0.25cm} & \hspace{-0.25cm} $\Gamma$ \hspace{-0.25cm} & \hspace{-0.25cm} $\log\xi$ \hspace{-0.25cm} & \hspace{-0.25cm} $A_{\rm Fe}$ \hspace{-0.25cm} & \hspace{-0.25cm} $\log N_{\rm e}$ \hspace{-0.25cm} & \hspace{-0.25cm} $E_{\rm cut}$ \hspace{-0.25cm} & \hspace{-0.25cm} $kT_{\rm e}$ \hspace{-0.25cm} & \hspace{-0.25cm} $l$ \hspace{-0.25cm} & \hspace{-0.25cm} $R_{\rm f}$ \hspace{-0.25cm} & \hspace{-0.1cm} \verb7defpar_type7 \hspace{-0.1cm} & \hspace{-0.1cm} \verb7defpar_value7 \hspace{-0.1cm} \\

\hline

{\sc relline\_nk} & $\surd$ & $\surd$ & $\surd$ & $\surd$ & $\times$ & $\surd$ & $\surd$ & $\surd$ & $\surd$ & $\surd$ & $\times$ & $\times$ & $\times$ & $15$ & $\times$ & $\times$ & $\surd$ & $\times$ & $\surd$ & $\surd$ \\ 

{\sc relconv\_nk} & $\times$ & $\surd$ & $\surd$ & $\surd$ & $\times$ & $\surd$ & $\surd$ & $\surd$ & $\surd$ & $\times$ & $\times$ & $\times$ & $\times$ & $15$ & $\times$ & $\times$ & $\surd$ & $\times$ & $\surd$ & $\surd$ \\ 

{\sc relxill\_nk} & $\times$ & $\surd$ & $\surd$ & $\surd$ & $\times$ & $\surd$ & $\surd$ & $\surd$ & $\surd$ & $\surd$ & $\surd$ & $\surd$ & $\surd$ & $15$ & $\surd$ & $\times$ & $\times$ & $\surd$ & $\surd$ & $\surd$ \\ 

{\sc relxillCp\_nk} & $\times$ & $\surd$ & $\surd$ & $\surd$ & $\times$ & $\surd$ & $\surd$ & $\surd$ & $\surd$ & $\surd$ & $\surd$ & $\surd$ & $\surd$ & $15$ & $\times$ & $\surd$ & $\times$ & $\surd$ & $\surd$ & $\surd$ \\ 

{\sc relxillD\_nk} & $\times$ & $\surd$ & $\surd$ & $\surd$ & $\times$ & $\surd$ & $\surd$ & $\surd$ & $\surd$ & $\surd$ & $\surd$ & $\surd$ & $\surd$ & $\surd$ & $300$ & $\times$ & $\times$ & $\surd$ & $\surd$ & $\surd$ \\ 

{\sc rellinelp\_nk} & $\surd$ & $\times$ & $\times$ & $\times$ & $\surd$ & $\surd$ & $\surd$ & $\surd$ & $\surd$ & $\surd$ & $\surd$ & $\times$ & $\times$ & $15$ & $\times$ & $\times$ & $\surd$ & $\times$ & $\surd$ & $\surd$ \\ 

{\sc relxilllp\_nk} & $\times$ & $\times$ & $\times$ & $\times$ & $\surd$ & $\surd$ & $\surd$ & $\surd$ & $\surd$ & $\surd$ & $\surd$ & $\surd$ & $\surd$ & $15$ & $\surd$ & $\times$ & $\times$ & $\surd$ & $\surd$ & $\surd$ \\ 

{\sc relxilllpCp\_nk} & $\times$ & $\times$ & $\times$ & $\times$ & $\surd$ & $\surd$ & $\surd$ & $\surd$ & $\surd$ & $\surd$ & $\surd$ & $\surd$ & $\surd$ & $15$ & $\times$ & $\surd$ & $\times$ & $\surd$ & $\surd$ & $\surd$ \\ 

{\sc relxilllpD\_nk} & $\times$ & $\times$ & $\times$ & $\times$ & $\surd$ & $\surd$ & $\surd$ & $\surd$ & $\surd$ & $\surd$ & $\surd$ & $\surd$ & $\surd$ & $\surd$ & $300$ & $\times$ & $\times$ & $\surd$ & $\surd$ & $\surd$ \\ 

\hline\hline
\end{tabular}
\vspace{0.2cm}
\caption{\label{tab:flavors} List of the available models and the parameters of each model. $\surd$ means the parameter is part of the model and $\times$ means it is not. }
\end{table*}
%


\vspace{0.5cm}

\section{Conclusion}
\label{sec:concs}

We have presented the public release version of \textsc{relxill\_nk}, an extension of the relativistic X-ray reflection model \textsc{relxill} to include non-Kerr BHs. We have also presented the new model \textsc{relxilllp\_nk}, a non-Kerr extension of \textsc{relxilllp} where the corona is assumed to be an isotropically emitting point source at some height along the spin axis of the BH. We have shown that the error introduced by our general relativistic ray-tracing method does not introduce significant error as compared with the current observational error present in BH reflection spectrum observations. Finally, we compare the relativistic iron line and reflection spectrum in both the standard and lamppost configurations for different values of the deformation parameters in the Johannsen spacetime.

There are still some improvements that can, and are planned, to be made to the \textsc{relxill\_nk} model. As noted in Sec.~\ref{sec:comp}, while the accuracy of \textsc{relxill\_nk} as compared with \textsc{relxill} is within about 1-2\% for inclination angles up to $70\degree$, the error increases significantly for inclination angles $\iota\gtrsim75\degree$. Generally, this is not a problem as most X-ray reflection spectrum observations are from systems with inclination angles below $75\degree$, it would be good to have a model that is complete and accurate across the full range of parameters. Another improvement that is important for upcoming X-ray telescopes is to improve the overall accuracy of \textsc{relxill\_nk}. While current telescopes lead to BH spin estimates with errors of about 10\%, future telescopes such as eXTP~\citep{2018arXiv181204020Z} are predicted to reduce the error to about 1\%. In this case, the 1-2\% numerical error seen in \textsc{relxill\_nk} would significantly influence the data analysis of observations and the spin estimates. The goal is to reduce the numerical error by about an order of magnitude so that it does not significantly impact the data analysis.

\acknowledgements

We thank Alejandro Cardenas-Avendano for earlier collaboration on the subject of this paper. This work was supported by the Innovation Program of the Shanghai Municipal Education Commission, Grant No.~2019-01-07-00-07-E00035, the National Natural Science Foundation of China (NSFC), Grant No.~U1531117, and Fudan University, Grant No.~IDH1512060. A.B.A. also acknowledges the support from the Shanghai Government Scholarship (SGS). J.A.G. acknowledges support from the Alexander von Humboldt Foundation. S.N. acknowledges support from the Excellence Initiative at Eberhard-Karls Universit\"at T\"ubingen.


\end{document}